\documentclass[aps,prd,twocolumn,showpacs]{revtex4}
\usepackage{graphicx}

\begin{document}

\title{Fast Cosmological Parameter Estimation from  
Microwave Background Temperature and Polarization Power Spectra}
\author{Raul Jimenez}
 \email{raulj@physics.upenn.edu}
\author{Licia Verde}
 \email{lverde@physics.upenn.edu}
\affiliation{
Dept. of Physics and Astronomy, University of Pennsylvania, 209 South 33rd Street, Philadelphia, 
PA-19104, USA}
\author{Hiranya Peiris}
 \email{hiranya@astro.princeton.edu}
\affiliation{
Dept. of Astrophysical Sciences, Princeton University, Princeton, NJ-08544, USA}
\author{Arthur Kosowsky}
 \email{kosowsky@physics.rutgers.edu}
\affiliation{
Dept. of Physics and Astronomy, Rutgers University, 136 Frelinghuysen Road, Piscataway, NJ-08854, USA}

\date{\today}

\begin{abstract}
  We improve the algorithm of Kosowsky, Milosavljevic, and Jimenez (2002)
  for computing power spectra of the cosmic microwave background.
  The present algorithm computes not only
  the temperature power spectrum but also the E-mode
  polarization and the temperature-polarization cross
  power spectra, providing the accuracy required for current 
  cosmological parameter estimation. We refine the optimum set of
  cosmological parameters for computing the
  power spectra as perturbations around a fiducial model, leading to
  an accuracy better than 0.5\% for the temperature power spectrum throughout
  the region of parameter space within {\sl WMAP}'s first-year 
  3$\sigma$ confidence region. This accuracy is comparable to the
  difference between the widely-used CMBFAST code (Seljak and Zaldarriaga 1996)
  and Boltzmann codes. Our algorithm (CMBwarp) makes possible a
  full exploration of the likelihood region for eight
  cosmological parameters 
  {\it in about one hour on a laptop computer}.
  We provide the code to compute power spectra as well as the Markov
  Chain Monte Carlo algorithm for cosmological parameters estimation
  at {\tt http://www.physics.upenn.edu/$\sim$raulj/CMBwarp}.
\end{abstract}


\maketitle

\section{Introduction}

The recent Wilkinson Microwave Anisotropy Probe ({\sl WMAP}) first-year data
release has shown that, despite advances in our ability to compute the
power spectrum of microwave background anisotropies and increases in
computer power, cosmological parameter estimation is still relatively
slow and computationally expensive. For example, for the simplest
parameter space of flat, Lambda Cold Dark Matter (LCDM) models, the {\sl WMAP} team
needed about a week of supercomputer time (32 CPUs on an SGI Origin)
to determine the two parameters joint 2-$\sigma$ confidence contours in the parameter space
\cite{SpergelWMAP03}.  Allowing a non-flat geometry slows down the
analysis significantly and more than quadruples the time required to
draw the 2-$\sigma$ contours. The main bottleneck in the parameter
estimation process is the theoretical computation of the power
spectrum anisotropy for a very large number of models, scattered throughout
the parameter space. This is mostly done using publicly available codes such as
CMBFAST (\cite{sel96}) or CAMB (\cite{CAMB}).  To accelerate the
calculations, (DASH)\cite{KKS02} introduced a shortcut in the calculation of
the angular power spectra, based on analytical and semi-analytical
approximations.

Rigorous analysis of future microwave background and other
complementary data sets require recomputing the likelihood contours
for cosmological parameters each time new data becomes
available. Additionally for each new experiment, not one but many
likelihood analyses are required to explore the impacts of potential systematic
errors.  Not every researcher has access to months of supercomputer
time for these computations; a fast and reasonably accurate method to
compute confidence regions for cosmological parameters is needed.
Motivated by the enormous computational burden that a proper 3 to
4-$\sigma$ exploration of the likelihood hypersurface represents and
by the related need to understand more clearly the physical parameters
that control the shape of the power spectrum, Kosowsky, Milosavljevic
and Jimenez (KMJ) \cite{KMJ02} presented a new set of ``physical''
parameters formed from the usual cosmological parameters which have
nearly-orthogonal effects on the microwave background temperature
power spectrum. Thus, with this set of parameters, the efficiency of
Monte Carlo techniques for evaluating the likelihood region increases
significantly; these parameters have been used in the WMAP analysis
\cite{SpergelWMAP03,VerdeWMAP03} and then incorporated into Lewis'
suite of microwave background analysis codes \cite{lew02}. Subsequent
work in a similar vein constructed a set of normal parameters that
locally transformed the likelihood in parameter space into Gaussian
form\cite{CKK03}.

In the new KJM ``basis'', it is also possible to obtain an efficient
approximation of the temperature power spectrum in a region around the
fiducial model as a simple linear extrapolation from a fiducial model
power spectrum. This new technique makes possible to compute
approximate power spectra, several orders of magnitude faster than any
previous methods, and was extensively used by the {\sl WMAP}
collaboration \cite{VerdeWMAP03} in the preliminary analysis of the
data.

Despite a significant increase in computational speed and physical
insight afforded by the KMJ parameters, some remaining technical
issues prevented a direct application of the KMJ power spectrum fits
to real data.  In particular: (a) the parameters and the linear fits
were only computed for the temperature power spectrum and not for the
polarization; (b) the fiducial model used by KMJ had an optical depth
to the last scattering surface which was too small ($\tau=0.06$)
compared with the {\sl WMAP} best-fit model ($\tau=0.17$); (c) for a
fiducial model with higher $\tau$ some covariance remained between the
slope of the power spectrum $n$, the amplitude of the fluctuations
$\cal S$ and the optical depth parameter ${\cal Z}= \exp (-2 \tau)$
which suggested that the KMJ parameters ${\cal Z}$, ${\cal S}$ and $n$
were not completely orthogonal; (d) errors in the power spectrum
estimation became larger than 1\% in some relevant regions of the
parameter space.

Here we present a refined version of the KMJ algorithm, CMBwarp, that
solves these problems. We extend the method to include polarization
spectra, make all parameters nearly orthogonal, use a fiducial model
given by the best fit model to {\sl WMAPext} data\cite{SpergelWMAP03},
and obtain the power spectra using polynomial fits for improved
accuracy over linear extrapolations.  Throughout the region of
parameter space within {\sl WMAP}'s first-year data 3-$\sigma$
confidence region\footnote{3-$\sigma$ for each parameter, marginalized
over the other n-1 parameters.}, the accuracy is now comparable to the
difference between CMBFAST and other independent Boltzmann codes.  We
make CMBwarp publicly available from {\tt
http:/www.physics.upenn.edu/$\sim$raulj/CMBwarp}; in addition, we also
release accompanying Markov Chain code that can be used to perform
parameter estimation.

\section{Cosmological Parameters and Physical Quantities}
\label{sec:parameters}

We first review the KMJ choice of physical parameters and describe the
slightly modified set of parameters presented here which yield an
orthogonal basis for temperature and polarization spectra.  Our model
space is the standard class of inflation-like cosmological models,
specified by five parameters determining the background homogeneous
spacetime (matter density $\Omega_{\rm mat}$, radiation density
$\Omega_{\rm rad}$, vacuum energy density $\Omega_\Lambda$, baryon
density $\Omega_b$, and Hubble parameter $h$), two parameters
determining the spectrum of primordial scalar perturbations (scalar
amplitude ${\cal S}$ and power law index $n$), and a single parameter
$\tau$ describing the total optical depth since reionization.  For the
present we postpone additional complications such as tensors, massive
neutrinos or a varying vacuum equation of state. Throughout this
paper, we use a fiducial cosmological model corresponding to WMAP's
best fit LCDM model with parameters: $h=0.73$, $\Omega_m=0.27$,
$\Omega_b=0.045$, $\Omega_V=0.69$, $n=0.99$, $\tau=0.166$, ${\cal
S}=0.88$. As it will be clear below, the choice for ${\cal S}$ is
convention-dependent. In our convention ${\cal S}=0.88$ which
corresponds to $\sigma_8=0.929$ in the fiducial model. We will
construct fast and accurate approximations to the microwave background
power spectra in some region of parameter space surrounding this
fiducial model.

\subsection{Physical parameters for the Temperature power spectrum}

The five parameters describing the background cosmology induce complex
dependencies in the microwave background power spectra through multiple
physical effects; therefore their effects on the power spectra are not
orthogonal.  A characteristic scale in the power spectrum is the
angular scale of the first acoustic peak. It is advantageous to choose
this angular scale as a parameter which can be varied independently of
any other parameters.  This angular scale is in turn determined by the
ratio of the sound horizon at last scattering (which determines the
physical wavelength of the acoustic waves) to the angular diameter
distance to the surface of last scattering (which determines the
apparent angular size of this yardstick).  KMJ used this quantity as
the first of the following set of orthogonal {\em physical} parameters.  The
analytic theory underlying any such choice of physical parameters has
been worked out in detail (see \cite{hs95a,hs95b,hs96,hw96}, and also
\cite{wei01a,wei01b}).

The characteristic angular scale is

\begin{equation}
{\cal A}\equiv {r_s(a_*) \over D_A(a_*)}
\label{parameterA}
\end{equation}

where $a_*$ is the scale factor at recombination,
the sound horizon is given by

\begin{eqnarray}
r_s(a) = {a\over H_0\sqrt{3}} \times \makebox[6cm] \nonumber \\
\!\!\!\!\!\!\int_0^a \!\!{dx \over
\left\{\!\left(1\! +\! {3\Omega_b\over 4\Omega_{\rm rad}}x\right)\!
\left[(1-\Omega)x^2 + \Omega_\Lambda x^4 \!+\! \Omega_{\rm m} x\! +\! \Omega_{\rm
rad}\right]\right\} ^{1/2}}
\label{rsa-integral}
\end{eqnarray}
%
and the angular diameter distance in a standard FRW spacetime by
\begin{equation}
D_A(a) = aH_0^{-1}|\Omega -1|^{-1/2}S_k(r),
\label{daa}
\end{equation}
with 
\begin{equation}
S_k(r) = \cases{\sin{r},& $\Omega > 1$;\cr
                r,&  $\Omega = 1$;\cr
                \sinh{r},& $\Omega < 1$;}
\label{Skdef}
\end{equation}

see, e.g.,  \cite{pea99}.
The other physical parameters employed by KMJ are:

\begin{eqnarray}
{\cal B} &\equiv& \Omega_b h^2,\cr
{\cal V} &\equiv& \Omega_\Lambda h^2,\cr 
{\cal R} &\equiv& {a_*\Omega_{\rm mat} \over \Omega_{\rm rad}},\cr
{\cal M} &\equiv& \left(\Omega_{\rm mat}^2 + a_*^{-2}\Omega_{\rm rad}^2
\right)^{1/2} h^2;
\label{BVRMdef}
\end{eqnarray}
$\cal B$ is the physical baryon density determining the baryon driving
effect on the acoustic oscillations \cite{hs95a,hw96} (which modifies
the relative odd and even acoustic peak heights); $\cal R$ is the
matter-radiation density ratio at recombination; $\cal V$ determines
the late-time Integrated Sachs-Wolfe effect arising from a late
vacuum-dominated phase, but otherwise varying ${\cal V}$ while holding
the other parameters fixed represents a nearly exact degeneracy
(sometimes called the ``geometrical degeneracy''); $\cal M$ couples
only to other small physical effects and is an approximate degeneracy
direction. This choice of parameters is not unique, but this specific
set is particularly useful and easy to interpret.

Given values for $\cal A$, $\cal B$, $\cal V$, $\cal R$, and $\cal M$,
they can be inverted to the corresponding cosmological parameters
by rewriting the definition of $\cal A$ in terms of $\cal B$,
$\cal V$, $\cal R$, $\cal M$, and $h$, then searching in $h$ until
the desired value for $\cal A$ is obtained; $\Omega_b$ and $\Omega_\Lambda$ 
then follow immediately, while $\Omega_{\rm mat}$ and $\Omega_{\rm rad}$
can be obtained with a few iterations to determine a precise value
for $a_*$. 
The other cosmological parameters which affect the microwave
background power spectrum are the optical depth to reionization and
the amplitude and power law index of the scalar perturbation power
spectrum.  These turn out to be crucial parameters in order to make an
approximate method like the one described here to work.
For reionization, KMJ used the physical parameter 
\begin{equation}
{\cal Z}\equiv e^{-2\tau},
\label{Zdef}
\end{equation}
the factor by which the microwave background anisotropies on small
scales are damped due to Compton scattering by free electrons after
the Universe is reionized. This damping occurs for all scales smaller
than the horizon size at reionization.

The primordial power spectrum of density perturbations is generally
taken to be a power law, $P(k) \propto k^n$, with $n=1$ corresponding
to the scale-invariant Harrison-Zel'dovich spectrum.
Making the approximation that $k$ and $l$ have a direct
correspondence, the effect of $n$ on the microwave background
power spectrum can be modeled as
\begin{equation}
C_{\ell}(n) = C_{\ell}(n_0) \left({\ell}\over {\ell}_0\right)^{n-n_0},
\label{n_effect}
\end{equation}
which is a good approximation for power law power spectra. A first-order
deviation from a power law, characterized by $\alpha = d\ln n/d\ln k$
\cite{kos95}, can be represented in a similar way.  Note that since
the dependence is exponential, a linear extrapolation is never a good
approximation over the entire interesting range $2<\ell<3000$.  The
choice of $\ell_0$ is somewhat arbitrary: changing $\ell_0$ simply gives a different
overall normalization.
In this work we choose $\ell_0=550$ as the best pivot point to
reproduce the CMBFAST power spectra.

These parameters have nearly orthogonal effects on the power spectra,
so the effect of varying a particular parameter can be considered
independently from the other parameters, within a reasonably large
region of parameter space.  This property can be exploited to create
simple and efficient approximations to the microwave background power
spectra, based on a few computed power spectra.  The KMJ paper chose a
fiducial cosmological model described by a set of parameters ${\bf
s}_0$ and computed its power spectrum $C_{\ell}({\bf s}_0)$ using available
codes.  The effects of parameter ${\cal A}$ were modeled as a simple
rescaling of the angular multipole $\ell$ and the effects of ${\cal
S}$ as a overall normalization factor, the effects of $n$ as in
equation \ref{n_effect} above.  Then for each of the remaining
parameters $s_i$, a linear fit was used to describe the variation of
$C_{\ell}({\bf s}_0, s_i)$ as $s_i$ moved away from its value in the
fiducial model.

This set of numerical fits gives approximations to
the power spectrum which are remarkably good in some region of
parameter space surrounding the fiducial model. However, the linear
extrapolations are not good enough to compute the power spectrum to
the accuracy of the CMBFAST code used to compute the fiducial model
and the dependencies on the individual parameters. Here we improve on a
simple linear extrapolation by performing polynomial fits (to the 4th
order) to the computed power spectra dependencies. 

\begin{figure}[ht]
\includegraphics[scale=0.52]{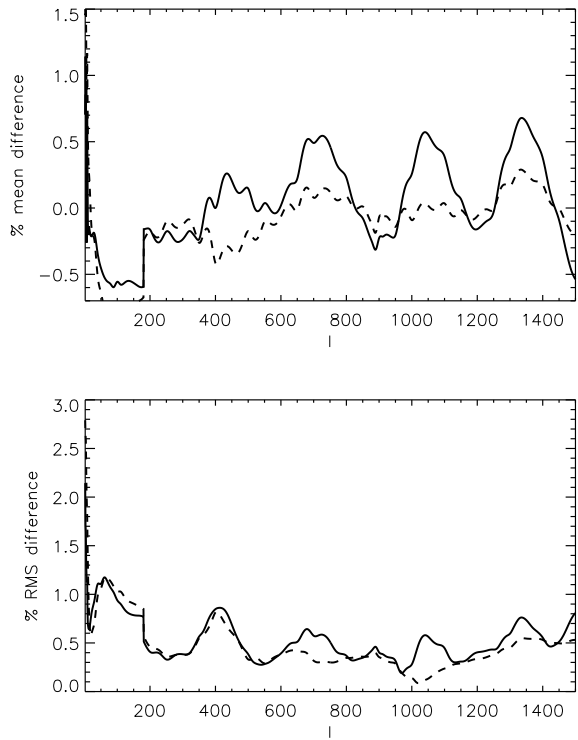}
\caption{Temperature power spectrum comparison between CMBFAST and
CMBwarp for $10^4$ models extracted randomly from a MCMC for {\sl
WMAP} first year data.  The solid line is for flat models while the
dashed line is for universes with arbitrary geometry. The top panel
shows the percent mean difference averaged from $l=2$ to $l=1500$,
while the bottom panel shows the percent RMS difference. Note that the
agreement is at the 0.5\% level or better.}
\label{fig:ttrms}
\end{figure}

\begin{figure}[ht]
\includegraphics[scale=0.52]{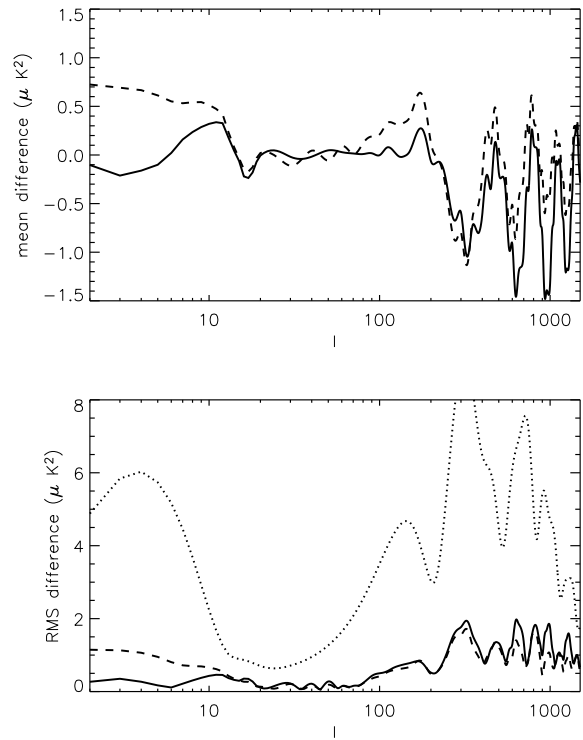}
\caption{Same as fig.~\ref{fig:ttrms} but for the TE power spectrum. 
Since TE 
crosses zero at several points, the vertical axis is plotted in $\mu
K^2$. The dotted line in the bottom panel shows the cosmic variance
error for the fiducial model.}
\label{fig:terms}
\end{figure}

\begin{figure*}[ht]
\includegraphics[scale=0.67]{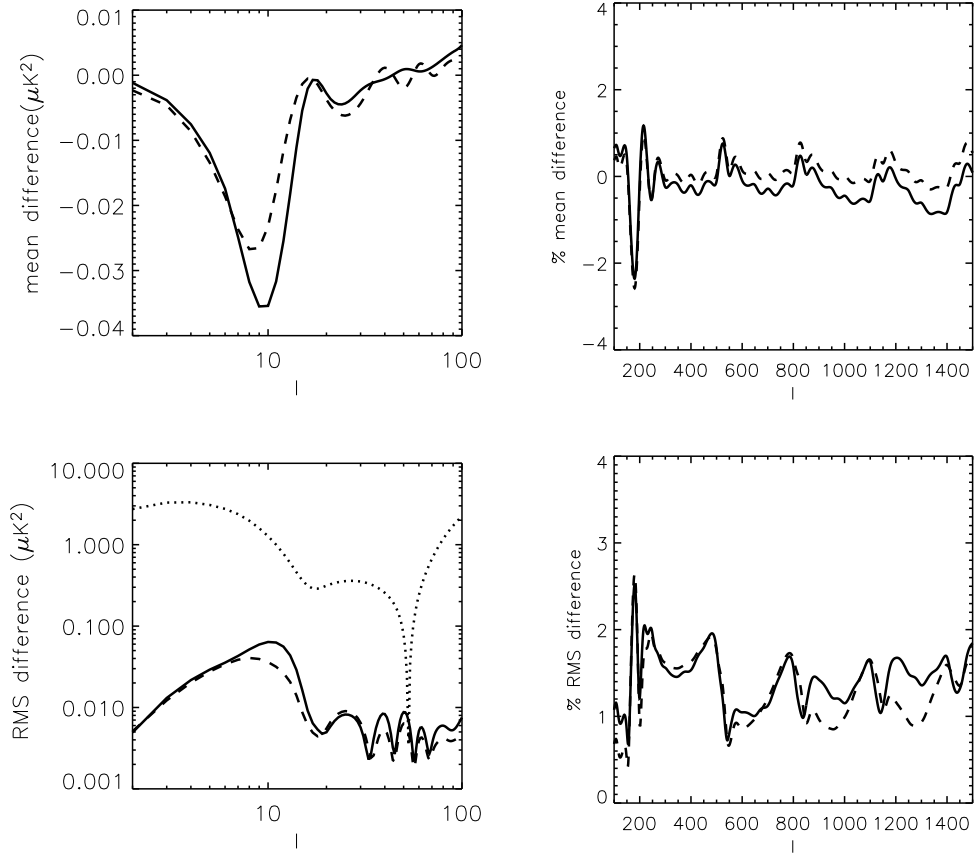}
\caption{Same as fig~\ref{fig:ttrms} but for the EE power
spectrum. For ${\ell} < 100$ we show the residual comparison (as in
fig.~\ref{fig:terms}) since the power spectrum is close to zero for
few multipoles. The dotted line shows the cosmic variance error for
the fiducial model. Note that the agreement with CMBFAST is at the
1-2\% level and always below the cosmic variance.}
\label{fig:eerms}
\end{figure*}

\begin{figure*}[ht]
\includegraphics[scale=0.83]{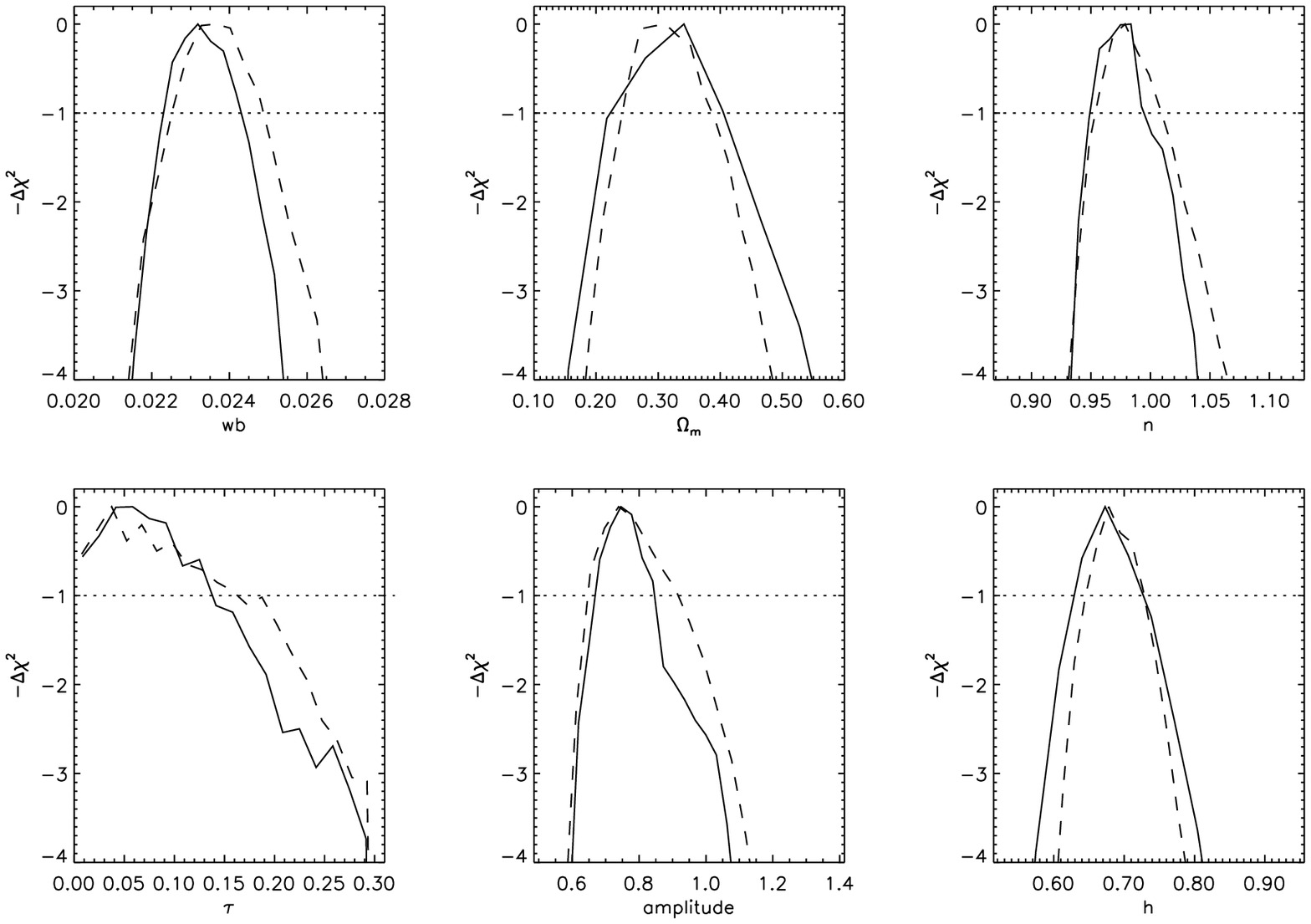}
\caption{Marginalized 1-D likelihood contours for cosmological
parameters recovered from the 1-year {\sl WMAP} TT dataset with
CMBFAST(dashed line) and those recovered with CMBwarp (solid line). It
is apparent that there is no bias in the recovered parameters with
CMBwarp. Even for $\tau$, which is virtually unconstrained, the
recovered CMBwarp likelihood shape is very similar to that of
CMBFAST. The dotted line is the 1-$\sigma$ level.}
\label{fig:wmaptt1d}
\end{figure*}

Unfortunately, the KMJ choice of physical parameters for $\cal Z$ and $n$
does not work if the fiducial model has a high $\tau$.
Even small deviations from the fiducial model result in
strong covariances between the two parameters (see further discussion
in Sec.~IV). Fortunately, a simple technique breaks the
degeneracy between these two parameters: we use $\cal Z$ as a
multiplicative factor to the whole power spectrum parameter for ${\ell} >
180$. For ${\ell} < 180$ the shape of the spectrum does not depend on $Z$
but only on $n$ as follows:

\begin{equation}
C_{\ell}(n)=C_{\ell}(n_0) \times ({\ell}/180)^{0.8(n-n_0)}
\label{n_corr}
\end{equation}

where the $0$ subscript denotes the fiducial value.  We then fix the
normalization for $\ell< 180$ by imposing the condition that the
spectrum must be continuous at $\ell=180$. For the amplitude $\cal S$ we 
use the same definition as in KMJ.

In the rest of the paper, we refer to the parameters $\cal A$, $\cal
B$, $\cal V$, $\cal R$, $\cal M$, $\cal S$, and $\cal Z$ as ``physical
parameters'', as opposed to the usual ``cosmological parameters''
$\Omega_b$, $\Omega_{\rm m}$, $\Omega_{\rm rad}$, $\Omega_{\rm vac}$,
$h$, $z_r$, and a quadrupole-based normalization. The set of
approximation techniques for the physical parameters, comprising
polynomial fits for ${\cal A}$, ${\cal B}$, ${\cal V}$, ${\cal R}$,
and ${\cal M}$, a multiplicative ${\cal Z}$ dependence, and an
approximate $n$ dependence given by Eqs.~(\ref{n_effect}) and
(\ref{n_corr}) provide an extremely simple, fast, and accurate method
of computing the temperature power spectrum of the microwave
background. Slight additional modifications at low $\ell$ values
described below provide excellent fits to the E-polarization and the
TE cross-power. All of these approximations taken as a whole comprise
the CMBwarp algorithm.

\subsection{Modeling the polarization power spectra}

For the cross power spectrum between the temperature and the E-mode
polarization, no modification is done for $\ell < 180$ and the whole
spectrum is modeled using the the same prescription for the
temperature for $\ell > 180$ but instead of the multiplicative factor
${\cal Z}$ we use the corresponding polynomial fit for ${\cal Z}$.
This models accurately the TE power spectrum neglecting secondary
effects. However, reionization of the neutral intergalactic medium by
the first generation of stars re-scatters the CMB photons, producing
the so called ``reionization bump'' at large angular scales. We model
this effect using the following fitting formula for $\ell < 16$,
inspired by \cite{Kaplinghat+03}:

\begin{equation}
C^{TE}_{\ell}=C^{TE\; {\rm fid}}_{\ell'}   \frac{1-{\cal Z}^{0.5}}{1-{{\cal Z}_0}^{0.5}} \left( {\frac{\log {{\cal Z}_0}}{\log {\cal Z}}} \right ) ^{0.21} (2/100)^{(n-1)}
\end{equation}

where $C^{TE{\rm fid}}_{\ell'}$  is the TE $C_{\ell'}$ value of the
fiducial model for each $\ell'$, evaluated at
\begin{equation}
\ell'={\rm int}[(0.105({\cal Z}/{\cal Z}_0)^8+0.93)\ell].
\end{equation}

The subscript $0$ denotes the fiducial model and ${\rm int}(x)$
denotes the integer part of $x$; we smoothly interpolate in the
transition region. This approximation is accurate to a few percent.

\begin{figure*}[ht]
\includegraphics[scale=0.83]{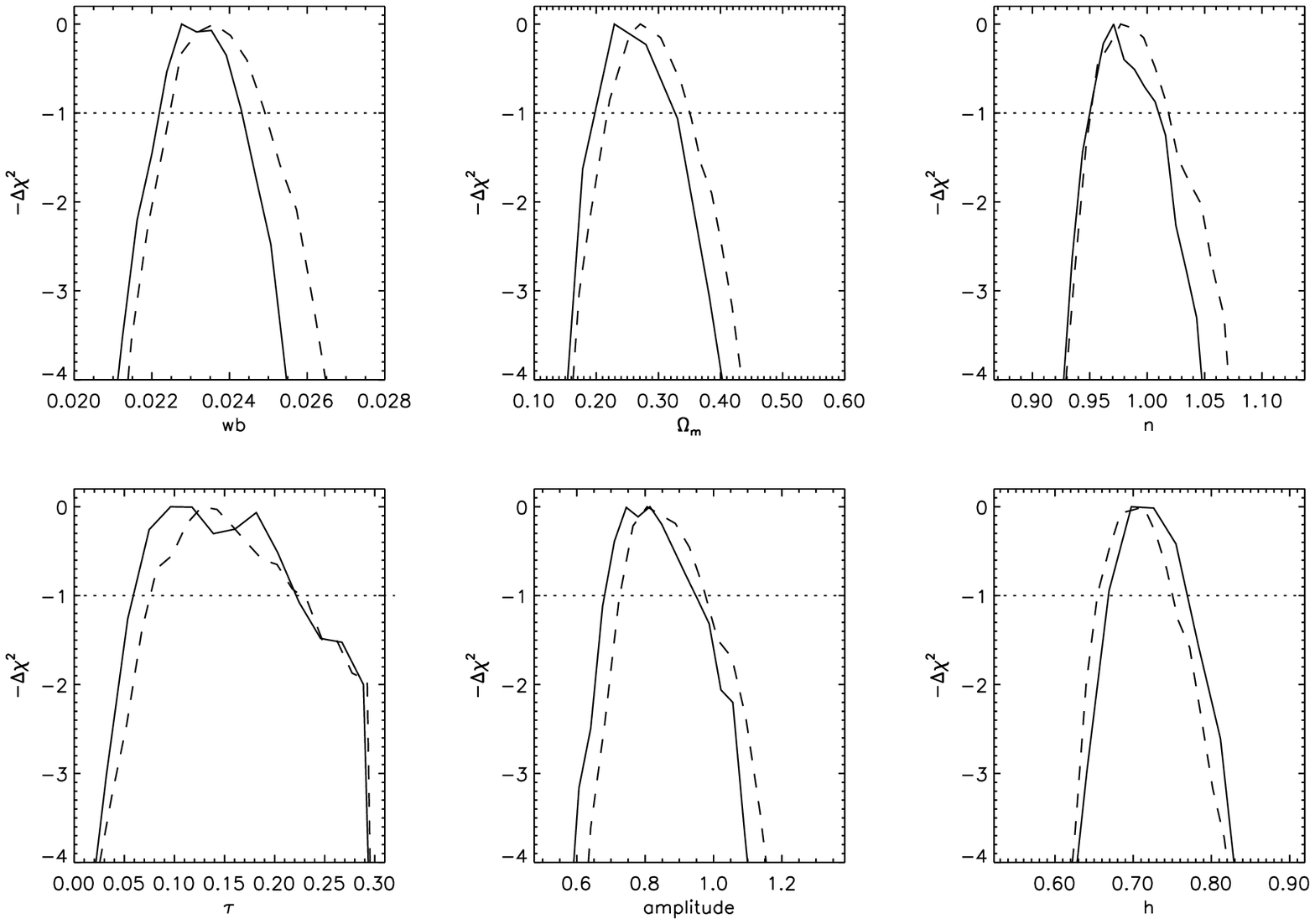}
\caption{Marginalized 1-D likelihood contours for cosmological
  parameters recovered from the 1-year {\sl WMAP} TT+TE and ACBAR TT
  datasets with CMBFAST(dashed line) and those recovered with CMBwarp
  (solid line). Note the good agreement between the two likelihoods
  for all parameters and the absence of a bias in both the best
  fitting cosmological parameters and the likelihood contours.}
\label{fig:wmapTTTE1d}
\end{figure*}

For the E-polarization power spectra, the power spectrum for all ${\ell}$'s is
modeled using the same prescription as for the temperature power
spectrum for $\ell > 180$ but instead of the multiplicative factor
${\cal Z}$ we use the corresponding polynomial fit for ${\cal Z}$.

\section{Performance of the Approximations}

The perturbative approach of CMBwarp greatly speeds up the calculation
of CMB power spectra since it requires only a few floating point
operations per multipole moment calculated. Computation time for a
single non-flat LCDM model is on the order of {\em 0.005 seconds} for
CMBwarp; about 100 seconds for CMBFAST on an Athlon 2400+
workstation. In other words, CMBwarp is about 20 thousand times
faster than CMBFAST. We have computed our numerical fits up to
${\ell}$ of 1500 and tested thoroughly the performance of the fit in
the same ${\ell}$ range. We have checked that the same type of
approximation also holds up to ${\ell}$ of 3500. Here we present a
direct application to data using {\sl WMAP} first year data and finer
angular scale experiments which probe scales up to ${\ell}$ of 1500.

For CMBwarp to be useful in the analysis of present and forthcoming
high-precision data sets, it must be not only fast but also accurate.
Two key questions are: (a) can the approximate approach reproduce,
with reasonable accuracy, CMBFAST power spectra? and (b) can it
recover unbiased and accurate cosmological parameters and confidence
regions in a likelihood analysis for a realistic data set?  We
demonstrate below that the answer is yes.

\subsection{Comparison between CMBFAST and CMBwarp} 

The first test we perform is to compare power spectra from CMBFAST and
CMBwarp for a large set of models.  For the comparison set, we select
models from converged Markov Chain Monte Carlos for the {\sl WMAP}
one-year data for both flat and non-flat LCDM models.  The chains are
run using CMBFAST (version 4.4 using the high precision option) and
the $C_{\ell}$ are saved; we select models with parameters within
(roughly) the $3 \sigma$ marginalized confidence level region for each
parameter. We then calculate power spectra for CMBwarp and compare the
two sets of $C_{\ell}$.  For TT, TE and EE power spectra we present
both the mean difference and the root-mean-square between CMBFAST and
CMBwarp, as a function of multipole $\ell$. The first quantity is an
estimate of the possible bias introduced by the perturbative
approximation, while the latter is an estimate of the scatter.  A
comparison between CMBFAST and Boltzmann codes \cite{SSWZ03} for one
cosmological model (the standard LCDM model) reveals differences at
the 0.1\% level in the temperature power spectrum and the 1\% level in
the polarization power spectrum.

Figure~\ref{fig:ttrms} shows this comparison for $10^4$ flat (solid
line) and $10^4$ open (dashed line) models for the temperature power
spectrum. The average difference is typically 0.3\%: of the same order of magnitude of 
the difference between CMBFAST and Boltzmann codes.  The RMS plot
shows not only that CMBwarp is unbiased, but that the RMS scatter is
only at the 0.5\% level. As we demonstrate below, this level of
accuracy is sufficient for accurate recovery of cosmological
parameters from current and future CMB temperature experiments.

The TE power spectrum crosses zero at several values of $\ell$, so a
fractional comparison is not possible. Instead we compute the residual
difference in $\mu$K$^2$ between the two models. To be more
specific we compute $\sqrt{(\sum(C_{\ell}({\rm CMBFAST})-C_{\ell}({\rm
CMBwarp}))^2)}/N_{\rm models}$ (RMS) and $\sum(C_{\ell}({\rm
CMBFAST})-C_{\ell}({\rm CMBwarp}))/N_{\rm models}$
(bias). Figure~\ref{fig:terms} shows the results. The discrepancy is
largest at large $l$, and is of the order of a 1-2\% percent. These
discrepancies are always small compared to the cosmic variance
$\sqrt{(C_\ell^{TT}C_\ell^{EE} + (C_\ell^{TE})^2)/(2\ell+1)}$, as
shown in the figure.

\begin{figure}[ht]
\includegraphics[scale=0.8]{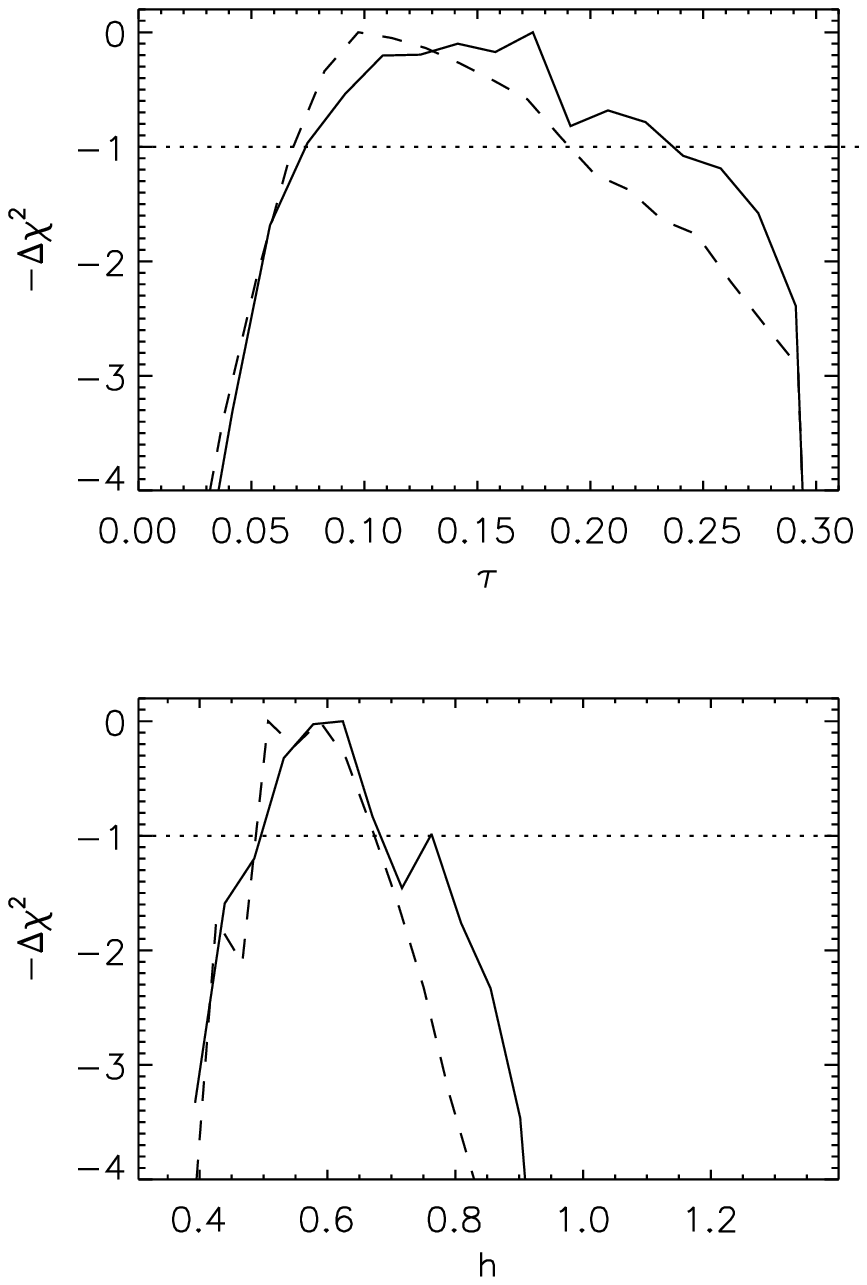}
\caption{Non-flat model 1-d likelihood for $h$ and $\tau$ using {\sl WMAP}
TT +TE. The dashed line is for CMBFAST while the solid line is for
CMBwarp.}
\label{fig:wmapopen}
\end{figure}

Finally, fig.~\ref{fig:eerms} shows the comparison between CMBFAST and
CMBwarp for the EE power spectrum. For $10 < {\ell} < 100$ the EE power spectrum
is very close to zero, thus for ${\ell} < 100$ we do a residual comparison as for TE. The
discrepancy with CMBFAST is only  few \% of a $\mu K^2$ and
significantly below cosmic variance. For ${\ell} > 100$ the EE power
spectrum is not close to zero and we show a fractional comparison. The
agreement with CMBFAST is at the 1-2\% level, similar
to that between CMBFAST and Boltzmann codes.

\subsection{CMBwarp MCMC and parameter estimation for  {\sl WMAP} 
first-year data}
\label{sec:errors}

We now turn our attention to the  determination of cosmological parameters
and their errors.  Given the satisfactory comparison between our
approximate power spectra and those from CMBFAST, we need to
demonstrate that when we determine cosmological parameters from real
measurements, CMBwarp yields the same parameter values and
confidence intervals as CMBFAST.

For this purpose we have calculated likelihood regions for two data
sets: first the {\sl WMAP} one-year temperature power spectrum data
\cite{HinshawWMAP03}, second the {\sl WMAP} one-year temperature and
TE power spectra \cite{KogutWMAP03} complemented by the ACBAR data points
\cite{ACBAR} extending to $l \simeq 1500$. The above tests are
performed for flat and open cosmologies. In the likelihood analysis we
use the full covariance matrix \cite{VerdeWMAP03} and we explore the
likelihood surface with a Markov Chain (see \cite{VerdeWMAP03} for
more details).  The procedure is effectively equivalent to that
illustrated in \cite{VerdeWMAP03}, but with CMBwarp replacing CMBFAST.
A Markov Chain with $10^5$ models is now computed in 1.5 hours on a
single Intel 3.2 GHz CPU, the computational bottleneck in the
calculation now being the evaluation of the likelihoods and not the
computation of the power spectra.

Figure~\ref{fig:wmaptt1d} shows the marginalized one-dimensional
likelihoods obtained from chains computed with CMBFAST (dashed line)
and CMBwarp (solid line), for {\sl WMAP} 1yr TT data only and for a flat
model.  The recovered best-fit parameters display no significant bias
(the parameter shift is always less that $0.25\sigma$), the 1-$\sigma$
error is recovered to within $15\%$ accuracy or better, and the
2-$\sigma$ error is recovered to 20\%.

Note that despite the fact that $\tau$ is effectively unconstrained
when only temperature data are used, CMBwarp still recovers the
same shape of the likelihood for $\tau$.

In the second test including the small-scale temperature measurements
and the {\sl WMAP} one-year TE data, the extra polarization data help to
greatly constrain the integrated optical depth to reionization. Once
again, chains of flat models generated with CMBFAST and CMBwarp
return virtually the same cosmological parameters as can be seen in
Fig.~\ref{fig:wmapTTTE1d}.  Finally, Fig.~\ref{fig:wmapopen} compares
the recovered values of $\tau$ and $h$ when the model is not
constrained to be flat.

\section{Discussion and Conclusions}
\label{sec:discussion}

The set of approximations described here, CMBwarp, comprise a
highly accurate and extremely fast method of approximating the
microwave background temperature and polarization power spectra in
standard inflation-like cosmological models for the range of
cosmological parameters which encompass the likelihood regions allowed
by the {\sl WMAP} first-year data. For 8 parameters, CMBwarp computes
each multipole moment with $\sim$ 50 floating-point
operations; it is unlikely that any significantly more efficient
computational scheme can be constructed.  Note that the computational
effort is the same for all multipole moments independent of $\ell$;
conventional codes require large increases in computational cost as
larger-$\ell$ multipoles are computed. This fact will become more
important in the coming years as experiments such as Planck,
ACT \cite{kosowsky03} and the South Pole Telescope map the microwave
background at high resolution.

CMBwarp is particularly well-suited to explorations of cosmological
parameter space using Markov Chain techniques. As cosmological data
continues to improve, joint analysis combining many different data
sets will continue. The technique allows easy incorporation of the
true microwave background constraints into joint analysis with a
minimum of computational expense. A useful extension of CMBwarp would
be a comparably fast and accurate computation of the transfer function
and matter power spectrum. The optimal set of ``physical'' parameters
for describing the matter power spectrum will not necessarily be the
same as for the microwave background; the crucial property of a set of
parameters is that they must have orthogonal effects on the matter
power spectrum. This will be addressed elsewhere. The speed and
flexibility of CMBwarp make possible analysis with many more
parameters which are not computationally feasible using standard
techniques: one example is estimation of the primordial power spectrum
using a number of parameters to describe it (see, e.g.,
\cite{mukherjee03}), taking into account the effects of the standard
cosmological parameters.

Markov Chain Monte Carlo explorations of cosmological parameter space have
now become common. The need for very fast, accurate, and portable
microwave background power spectrum computations in this context is
readily apparent. We hope CMBwarp proves to be a useful tool
for these analyses.

\begin{acknowledgments}
The work of RJ is partially supported by NSF grant AST-0206031. We
thank P.~G.~Castro and Andy Taylor for useful discussions. RJ and LV
thank the ``Centro de Investigaciones de Astronomia'' (CIDA) and the
University of Edinburgh for hospitality while part of this work was
carried out. AK is a Cottrell Scholar of the Research Corporation.
We acknowledge the use of the Legacy Archive for Microwave Background
Data Analysis (LAMBDA). Support for LAMBDA is provided by the NASA Office
of Space Science.
\end{acknowledgments}


\end{document}